\documentclass{vgtc}                          

\graphicspath{{figures/}{pictures/}{images/}{./}}

\usepackage{times}

\usepackage{tabu}
\usepackage{booktabs}
\usepackage{lipsum}
\usepackage{mwe}
\usepackage{svg}
\usepackage{tabularx}
\usepackage{xcolor}
\usepackage{booktabs}

\newcommand{\change}[1]{{#1}}
\usepackage{mathptmx}

\onlineid{1195}
\vgtccategory{Research}
\vgtcinsertpkg

\newif\ifnotes
\notestrue

\title{Debugging as Evidence-Driven Reasoning: Visualization Opportunities in Data-Intensive Programming}

\author{Yongbo Chen\thanks{e-mail: \{ychen88, yzhu27, rfaust1\}@tulane.edu}\\ %
        \scriptsize Tulane University %
\and Yan Zhu\footnotemark[1]\\ 
     \scriptsize Tulane University %
\and Rebecca Faust\footnotemark[1]\\\ %
     \scriptsize Tulane University}

\teaser{
  \centering
  \includegraphics[width=.95\linewidth]{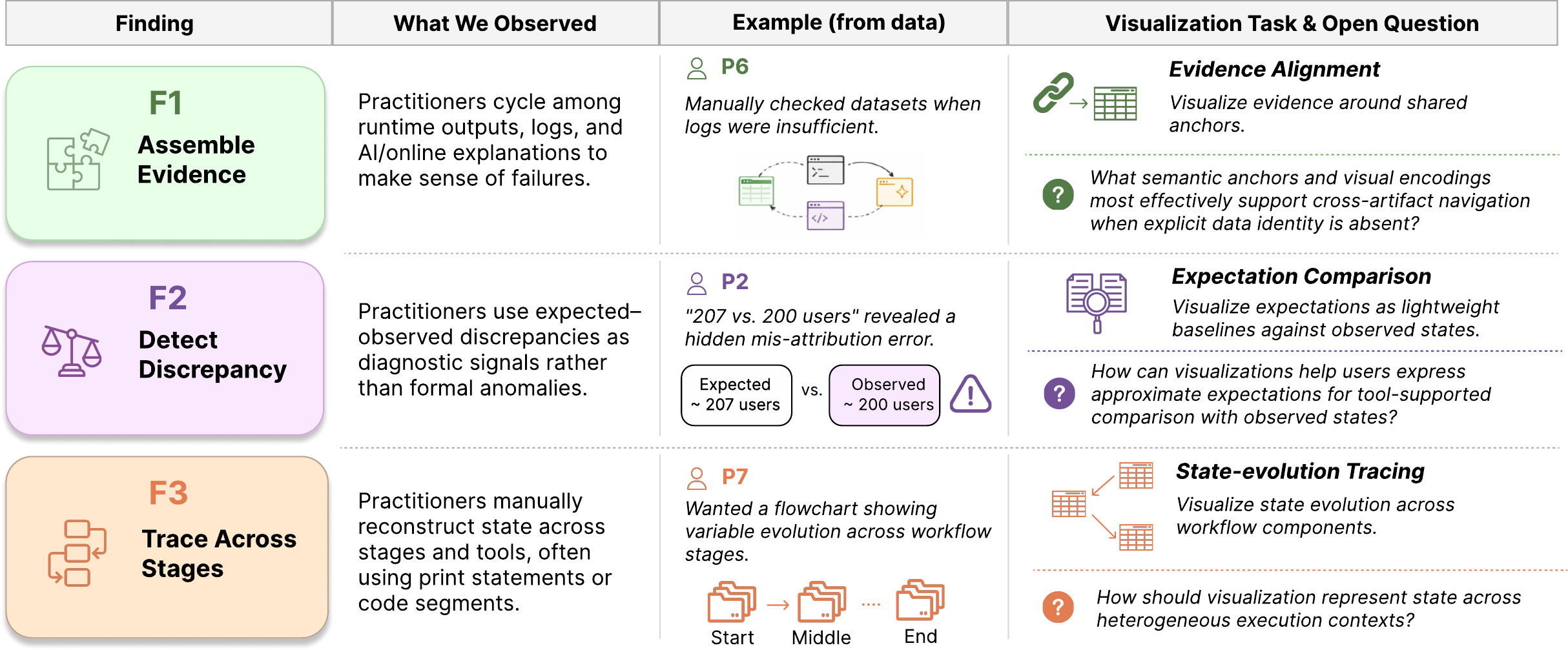}
  \vspace{-1em}
  \caption{Based on interviews with nine data-intensive participants, we identify three cross-cutting findings and map them to visualization opportunities: evidence alignment (F1), expectation-grounded comparison (F2), and state-evolution tracing (F3). Each row pairs a finding with an observation, an example, and the corresponding visualization requirement and open research question.}
  \label{fig:teaser}
}

\abstract{
    Visualization has been recognized as a valuable means of supporting debugging by externalizing runtime behavior that would otherwise remain hidden or scattered. 
    However, most visual debugging research has focused on traditional software development settings, leaving the distinct challenges of data-intensive workflows largely uncharacterized. 
    To build visual debugging support for these settings, we first need to characterize how practitioners debug in these settings and translate their challenges into concrete visualization opportunities. To this end, we conducted semi-structured interviews with nine participants from diverse data-intensive domains and analyzed the data using thematic analysis. Our analysis reveals three cross-cutting challenges: assembling fragmented evidence, detecting expected--observed discrepancies, and tracing state evolution across workflow components. 
    We distill these challenges into three concrete requirements that \change{current debuggers support only partially} but that visualization is well suited to address: cross-artifact evidence alignment, expectation-grounded comparison, and traceable state evolution. Together, these requirements \change{begin to characterize} a design space for future visual debugging research in data-intensive programming.
}


\begin{document}

\firstsection{Introduction}
\maketitle

Debugging is fundamentally a reasoning-intensive activity: developers must interpret program behavior, answer questions about execution, and judge whether observed outputs match their intent \cite{latoza2010developers, sillito2008asking}. Because this reasoning often depends on information that is invisible or scattered across runtime artifacts, visualization has long been recognized as a way to support debugging by externalizing program behavior into observable and inspectable forms \cite{faust2024anteater, ko2004designing}. 

However, most visual debugging research has focused on traditional software development settings, where debugging centers on locating and fixing isolated faults in code. 
\change{In contrast, the settings we study foreground data transformations, intermediate states, and domain-specific expectations across workflow components; we use \textit{data-intensive programming} to refer to this form of programming work, where debugging outcomes depend on these factors rather than on control flow or code structure alone. This framing includes computational notebooks as a prominent setting, as well as SQL-to-script workflows, non-interactive data-processing scripts, database-backed analyses, and dashboard-oriented pipelines. For instance, a pipeline that joins records in a database, transforms them in a notebook, and renders them in a dashboard may run without error yet still return an undesired result, such as a miscounted aggregation, which surfaces only when intermediate states across stages are compared against what the analyst expects. Debugging in these settings introduces challenges that visual debugging research has less directly addressed:}
multi-stage transformations, heterogeneous toolchains (e.g., notebooks, databases, and dashboards), and limited visibility into intermediate states \cite{chattopadhyay2020s, ramasamy2023visualising}. Despite these challenges, we lack detailed empirical accounts of how data-intensive practitioners actually navigate them, and how visualization research should respond to its specific demands.

To address this gap, we conducted semi-structured interviews with nine data-intensive professionals from diverse fields and used thematic analysis to examine their debugging challenges and strategies, information-seeking practices, and expectations for tool support. Our analysis reveals recurring patterns in how practitioners assemble, compare, and reason about heterogeneous evidence during debugging. From these patterns, we derive three cross-cutting findings relevant to visualization design (summarized in Fig.~\ref{fig:teaser}): evidence fragmentation across tools and representations, discrepancy-driven reasoning between expected and observed states, and limited visibility into state evolution across workflow components. We distill these into three requirements for visual debugging support \change{that are not well addressed by current debuggers}: cross-artifact evidence alignment, expectation-grounded comparison, and traceable state evolution. These requirements \change{outline an initial design space} for future visual debugging research in data-intensive programming.

\section{Related Work}

\subsection{Debugging Cognition and Information Seeking}
Previous research on debugging cognition has framed debugging as a process of asking questions about program behavior and seeking relevant information during code changes~\cite{sillito2008asking}. For example, Whyline \cite{ko2004designing} reframes debugging from searching through code to explaining observable behavior by enabling developers to ask ``why did'' and ``why didn't'' questions.
Such question-asking and information-seeking activities reflect a broader cognitive pattern, where sensemaking models describe how people forage for, organize, and interpret information in order to form and refine explanations of complex situations \cite{pirolli2005sensemaking}. Empirical studies of software maintenance \change{and professional debugging practices} similarly show that developers spend substantial effort seeking, relating, and collecting information scattered across code, documentation, and runtime artifacts \cite{ko2006exploratory} \change{ and rely heavily on manual inspection and hypothesis testing, with limited adoption of advanced techniques such as automated fault localization~\cite{perscheid2017studying}.} 

This perspective has been applied to analyze the cognitive challenges faced by professional software engineers when tracing execution paths, determining reachability, and understanding behavior in code \cite{latoza2010developers, layman2013debugging}. It has also been extended to end-user programming: Grigoreanu et al. \cite{grigoreanu2012end} applied the sensemaking framework to end-user debugging in spreadsheets, finding that information foraging constituted a major portion of the debugging process and that unsuccessful debuggers tended to get stuck when attempting to synthesize evidence into coherent explanations.

However, these studies are primarily grounded in traditional codebases or relatively bounded programming environments, leaving data-intensive debugging practices in more complex workflow settings less well understood.

\subsection{Representation and Provenance Support}
In data-intensive programming environments, particularly computational notebooks, prior work has identified structural challenges that complicate code understanding, debugging, and iterative analysis. Chattopadhyay et al. \cite{chattopadhyay2020s} documented pain points across the notebook workflow, while Kery et al. \cite{kery2018story} and Ramasamy et al. \cite{ramasamy2023visualising} showed that data scientists struggle to manage analysis history, code variation, and non-linear execution.

To address some of these issues, efforts have sought to bridge different information representations within notebooks or enhance the visibility of provenance. \change{Some efforts seek to bridge code and interactive visualizations through a shared query representation~\cite{Wu2020B2} as well as support fluid movement between code and graphical interfaces within notebooks~\cite{kery2020mage}}
\change{Scully-Allison et al. \cite{scully2024design} examine design concerns for integrating scripting with interactive visualization in notebook environments.} Loops \cite{eckelt2025loops} uses provenance information to visualize notebook evolution and differences across versions in code, tables, and visual outputs. Other work \cite{macke2021nbsafety, stitz2016avocado} explored notebook state management and provenance-oriented workflow visualization, including support for hidden-state awareness and analytical workflow provenance. Outside notebook environments, SOMNUS \cite{Xiong2023somnus} provides provenance-oriented views of the creation and transformation semantics of data tables in wrangling scripts.

However, these efforts primarily target workflow comprehension, reproducibility, or representation management rather than the anomaly-driven reasoning process that characterizes debugging. How practitioners integrate and validate evidence from multiple sources during debugging therefore remains less well understood.

\subsection{Runtime Inspection and Data Comparison}
Existing debugging tools largely center on code-level logic, control flow, and breakpoint-based inspection. In data-intensive settings, recent systems have instead emphasized making runtime states and intermediate data more visible. For example, DeSQL \cite{haroon2024desql} decomposes DISC-backed SQL queries into inspectable fragments, allowing users to step through intermediate results. Anteater \cite{faust2024anteater} takes a visualization-first approach by tracing and visualizing program execution values in context. Texera \cite{DBLP:journals/pvldb/WangHNKALLDL24} further supports runtime interaction in workflow-based data analytics, allowing users to pause execution, inspect intermediate operator states, and modify processing logic. Other systems provide complementary forms of notebook data comparison and multi-level execution visualization \cite{hayatpur2023crosscode, wang2022dtil}.

While these systems primarily demonstrate the value of enhanced visibility and interactivity, they offer less insight into how practitioners actually use such runtime signals to assemble evidence, detect discrepancies, and reason about data state evolution during debugging. This gap motivates our focus on how practitioners use heterogeneous evidence during data-intensive debugging.

\section{Methodology}
We conducted a qualitative interview study to characterize how data-intensive practitioners reason about debugging and translate their challenges into concrete visualization opportunities, focusing on cross-domain patterns rather than prevalence estimation.

\subsection{Participants}
We recruited nine participants who regularly write, modify, or inspect code as part of data processing, analysis, or validation work. The study was reviewed and approved by our institution's Institutional Review Board. Participants were recruited through email outreach and professional networks as a convenience sample. As shown in Table~\ref{tab:participants}, they spanned a range of roles, domains, and programming languages, including Python, SQL, R, MATLAB, JavaScript, and SAS. Their self-reported programming experience ranged from beginner to advanced.

\subsection{Study Procedure}
Each semi-structured interview lasted 30 to 45 minutes, was conducted remotely via Zoom, and was audio-recorded. The interviews covered four areas: participants' background and programming experience, recent debugging experiences, information-seeking practices and tool use, and expectations for future tool support. We used a broad definition of debugging to include not only fixing errors, but also situations in which participants investigated unexpected behavior, judged whether outputs were trustworthy, or inspected intermediate states to explain why results diverged from expectations. 

\subsection{Data Analysis}
We analyzed the data using inductive thematic analysis \cite{braun2006using}. Two coders independently conducted open coding on the interview transcripts and met regularly to compare interpretations, discuss disagreements, and refine code definitions; disagreements were resolved through discussion until consensus was reached. Through iterative axial and selective coding \cite{saldana2013coding}, we grouped related codes into five broader themes --- debugging challenges, debugging strategies, information-seeking practices, attitudes toward existing support, and desired tool support. In the final stage of analysis, the team identified the three findings reported in this paper because each finding recurred across multiple themes and participants, and most directly informed visualization design opportunities rather than capturing domain- or tool-specific variation. \change{Concretely, we examined which code families recurred across themes and grouped those cross-cutting patterns into candidate findings; e.g., codes about consulting heterogeneous sources (under the information-seeking and strategy themes) converged into the evidence-assembly finding. The codebook and a theme-to-finding mapping documenting this synthesis are provided as supplemental material.}

\vspace{-0.4em}
\begin{table}[t]
\centering
\footnotesize
\renewcommand{\arraystretch}{1.08}
\setlength{\tabcolsep}{3pt}
\begin{tabularx}{\linewidth}{l l l X}
\toprule
ID & Role & Domain & Languages \\
\midrule
P1 & Research Assistant & Environ. Health & Python \\
P2 & Data Analyst & Marketing & Python, SQL, JS\\
P3 & PhD Student & Neuroscience & C++, Python, MATLAB \\
P4 & Undergraduate & Economics & Python, R, SQL \\
P5 & Postdoc Researcher & Neuroengineering & MATLAB, Python \\
P6 & Data Engineer & Data Infrastructure & Python, SQL \\
P7 & Business Analyst & Finance & Python, SQL \\
P8 & Systems Associate & Banking / Biostats & R, Python, SAS \\
P9 & PhD Student & Cognitive Neurosci. & Python, R, Bash \\
\bottomrule
\end{tabularx}
\vspace{0.2em}
\caption{Participant roles, domains, and programming languages.}
\vspace{-2.8em}
\label{tab:participants}
\end{table}

\section{Key Findings}
Based on the analytical approach described above, we identified three cross-cutting findings that characterize recurring patterns in participants' debugging practices. Together, these findings frame data-intensive debugging as an evidence-driven reasoning process and motivate specific opportunities for visualization support.

\subsection{F1: Debugging Requires Active Assembly of Fragmented Evidence}
\label{sec: F1}
\change{Participants’ debugging does not rely on a single authoritative
source; instead, they actively gather and cross-verify evidence from
multiple sources. } In practice, participants typically begin by examining runtime outputs and intermediate results to determine whether program behavior is normal (P1, P3, P4, P6--P9), and analyzing logs and error traces to pinpoint failures (P1--P8). When local signals are insufficient to explain anomalous behavior, they turn to AI tools (P2--P7, P9) or online resources (e.g., Stack Overflow) for explanations (P3, P5, P6, P7, P9), while using their domain knowledge of expected outcomes to assess the plausibility of these external explanations (P3, P4, P5, P8). Importantly, this process is rarely linear: participants cycled back and forth among runtime outputs, logs, and external explanations (P1, P2, P4, P6, P9), using each source to reinterpret or validate signals from the others.

However, relying on any single type of information source is often insufficient to pinpoint the problem independently. As one participant described, unclear error messages could force them to inspect underlying datasets directly:

\noindent\textit{``sometimes the error messages you get aren't very obvious \dots But sometimes I have to manually check, like, all the datasets, because the error message isn't so obvious.''} (P6)
\vspace{0.25em}

The core difficulty lies less in the absence of information than in assembly of  scattered signals into a coherent explanation. While debugging across multiple sources is a known phenomenon, our findings characterize a  specific data-intensive pattern: iterative cycling among heterogeneous evidence types to cross-validate signals.

\subsection{F2: Debugging Reasoning Is Triggered by Expected–Observed Discrepancies}
\label{sec: F2}
In many data-intensive debugging scenarios, debugging begins not with explicit error signals (such as crashes that halt program execution), but when participants notice a discrepancy between expected and observed output. This expected-vs-actual contrast serves as the primary trigger for their debugging reasoning. 

Specifically, based on prior experience, task expectations, and domain knowledge, participants form mental expectations regarding ``what normal output should look like'' (such as an approximate number of rows in a table, a reasonable range for a numerical value, or an expected pattern for an aggregation) (P2, P4, P8, P9). When the observed output does not match these expectations, they treat such a mismatch as a diagnostic signal to guide subsequent inspection and hypothesis testing (P3, P4, P5, P8). For example, P4 described this contrast in their workflow:

\vspace{0.3em}
\noindent\textit{``I generally know what the desired outcome should be. For example, with SQL, I have an idea of what the table is supposed to look like, or a rough idea of what the numbers should be. If something looks outrageous, then I know something's wrong''} (P4)
\vspace{0.25em}

However, in data-intensive workflows, many failures do not generate explicit error signals: programs may execute successfully and produce results that appear reasonable, yet are actually incorrect. As another participant noted, assessing whether a numerical result is ``normal'' is inherently difficult (P8), meaning plausible but incorrect outputs may be completely overlooked. Another example comes from P2, who described a case where a query returned 200 users instead of the expected 207, a small discrepancy that appeared negligible but turned out to reflect a systematic misattribution error across data channels.

In such cases, discrepancy detection relies heavily on human judgment and domain expertise, indicating a need for tools that make users' own task- or domain-grounded expectations  visible and comparable, rather than relying on generic anomaly detection.

\subsection{F3: Debugging is Hindered by Limited Visibility into State Evolution Across Workflow Components}
\label{sec: F3}
Participants often struggled to inspect how data and variables evolved across workflow stages, particularly when these stages spanned multiple tools or execution environments. Although final outputs in data-intensive workflows emerge through multiple stages of transformation (e.g., SQL queries, scripting environments, and downstream visualizations), the intermediate states produced along the way were often difficult to inspect systematically across these heterogeneous boundaries.

Lacking such visibility, participants are forced to rely on manual methods to reconstruct state evolution within each tool, such as inserting print statements (P3, P4, P5, P7, P8), checking intermediate values (P3, P5, P7, P9), or executing code in segments (P1, P2, P4, P5, P6, P7) to observe the output of each step. When debugging concerns extended across data sources or tool boundaries, participants resorted to manual cross-validation: P2, for example, manually grouped query results by channel and compared totals against another database to localize a misattribution issue. However, several participants (P3, P7, P8) noted that such manual strategies are inefficient and cognitively burdensome when dealing with complex workflows. Participants also articulated a need for visualizations that support tracking state across stages, as P7 expressed:

\vspace{0.3em}
\noindent\textit{``From my experience, I think a flowchart would be very useful ... That kind of visualization would make debugging much easier. If I spot that the variable changes in an unexpected way in the middle of the program, I can go back to the earlier part of the code and figure out what went wrong.''} (P7)
\vspace{0.25em}

This requirement indicates that participants expect more than just static feedback when errors occur; they also want to continuously observe the stepwise evolution of data and variables across workflow components, even when those components span different tools and execution environments. While the value of inspecting intermediate state is well-established for within-program debugging, our findings highlight a distinct concern: the need to follow state across heterogeneous execution contexts, where the same underlying data is represented differently as it moves between databases, scripting environments, and downstream visualizations. Such challenges motivate representations that make state evolution traceable across heterogeneous workflow components, rather than exposing only isolated snapshots or within-tool views.

\section{Discussion: Implications for future work}
Our findings collectively point toward visualization support structured around three complementary visual requirements that together enable evidence-driven reasoning during data-intensive debugging.

\paragraph{\textbf{Evidence Alignment}} Building on Sec. \ref{sec: F1}, we articulate the need as a concrete visualization requirement: visualizations should align evidence distributed across multiple sources around a shared anchor (i.e., a common reference point such as a workflow step, variable, or data entity), making it navigable from a single anomalous observation rather than presenting it side by side. 
For instance, when a user encounters an unexpected query result, a visualization could co-locate the runtime output, the relevant SQL query, the corresponding intermediate data snapshots, and any AI-generated explanations around the shared workflow step where the discrepancy emerged. 
The goal is to allow users to start from an anomalous observation and access all related debugging signals.

Prior visualization research has shown the value of coordinating heterogeneous information across sources through related views, visual fusion, and interactive linking \cite{kehrer2012visualization}. In our study, we find that data-intensive debugging poses an analogous need: users must coordinate evidence distributed across multiple artifacts and tools. This requirement also raises an open question for visual debugging research: when artifacts such as runtime values, logs, AI-generated explanations, and domain expectations lack a shared data identity, what anchors can most effectively support cross-artifact navigation, and how can visualization make these cross-artifact correspondences legible? This question is challenging because, as Sec.~\ref{sec: F1} showed, the signals that participants rely on are heterogeneous in both structure and modality, and their relationships are often semantic rather than explicitly encoded in shared data. Prior work on linked views has largely assumed identifiable correspondences within shared data \cite{roberts2007state}, while recent notebook systems have begun to link code, visualizations, and outputs in more structured environments \cite{lin2025interlink, Wu2020B2}. Future visual debugging tools should support coordination across evidence sources whose relationships are meaningful to users but not directly represented in the underlying data.

\paragraph{\textbf{Expectation-grounded Comparison}} Extending the discrepancy-driven pattern of Sec.~\ref{sec: F2}, we frame a second visualization requirement: visualizations should allow users to externalize the expectations they already rely on (e.g., plausible counts or  ranges) and visually compare observed states against them. Using P2's ``207 vs. 200 users'' example from Sec.~\ref{sec: F2} as illustration, a user could specify an expected count inline as a soft target, and the visualization could highlight the small but persistent deviation alongside the upstream transformation steps where user counts were attributed to channels, making an otherwise plausible-looking discrepancy visible at the comparison point.

Visualization has long been studied as a means of supporting comparison between complex data objects \cite{gleicher2011visual}. Our findings suggest that data-intensive debugging involves a related comparison problem, but with an important complication: one side of the comparison is often an implicit and approximate expectation held by the user. The design challenge, then, is not only how to compare two visible states, but rather
how can visualizations help users express approximate expectations in ways that remain lightweight for users yet interpretable enough for tool-supported comparison with observed states? This problem is difficult to resolve \change{because existing data-validation approaches assume expectations can be formalized in advance: assertions, constraints, or threshold-based checks \cite{deequ_repo, great_expectations_official, schelter2018automating} require a precise rule to be committed before the comparison, and generic anomaly detection flags statistical outliers rather than task-grounded ones. Such approaches are effective once an expectation is crisp enough to encode, but participants often had not yet reached that point: they reasoned from approximate, evolving intuitions --- P4, for instance, had only ``a rough idea of what the numbers should be.'' Expectation-grounded comparison is therefore not a replacement for assertions or validation, but support for the earlier, exploratory stage that precedes them---helping users externalize and inspect a fuzzy expectation until it becomes clear enough to act on, or to formalize as an assertion later.}
In this context, expectation-grounded comparison means treating such expectations as visible, revisable baselines for interpretation rather than forcing users to formalize them as precise constraints.

\paragraph{\textbf{State Evolution Tracing}} Sec.~\ref{sec: F3} points to a third visualization requirement. Visualization is particularly well suited to making changes in data state across stages perceptible at a glance, and our findings suggest that this affordance is important for debugging when practitioners need to trace how data attributes---including values, types, distributions, and schema changes---evolve across workflow components. Supporting this requirement would help users identify where behavior diverges and reconstruct the causal path, replacing the cumbersome manual practices described in Sec.~\ref{sec: F3}. For example, in a workflow that aggregates SQL query results in Python before rendering them in a dashboard, a visualization could expose stage-by-stage attribute changes and highlight where a metric first diverged from expected behavior.

A parallel question arises when data-intensive workflows span multiple execution contexts (e.g., SQL query $\rightarrow$ Python analysis $\rightarrow$ dashboard): How should visualizations represent state evolution across these heterogeneous boundaries? Prior work has made progress in revealing execution or provenance evolution in adjacent settings, including \change{object-mutation visualization \cite{schulz2016visually},} single-program trace visualization \cite{faust2024anteater}, notebook provenance \cite{eckelt2025loops}, and workflow provenance \cite{stitz2016avocado}. \change{More broadly, techniques such as program slicing \cite{weiser1984program}, breakpoint debugging \cite{moseler2020visual}, and formal verification help developers narrow dependencies, inspect execution state, or check specified properties, but operate within a single program over its control flow and code structure. Our participants' difficulty was different in kind: following \textit{data} state as it is transformed and re-represented across tool boundaries, where the relevant unit is the evolving data rather than a line of code.} Future visual debugging methods should treat state evolution as a representation that spans execution boundaries rather than as steps within a single program.

\section{Limitations}
\change{Our findings should be interpreted as qualitative design insights rather than prevalence estimates. Although participants spanned varied roles and environments, limited within-domain representation means that recurring patterns neither establish within-domain consistency nor generalize across communities. These insights are most transferable to data-state-centric, heterogeneous environments (e.g., notebooks, pipelines, databases, dashboards) and may not extend to embedded, interactive UI, concurrent, or purely algorithmic debugging contexts. Our reliance on retrospective self-reports is also subject to recall and rationalization bias. The three requirements are therefore design directions, not validated solutions; future work could combine in-situ and trace-based studies with prototype evaluations to test them at larger scales.}

\section{Conclusion}
Through interviews with nine data-intensive practitioners, we identified three cross-cutting challenges that characterize debugging as evidence-driven reasoning: assembling fragmented evidence, reasoning through expected--observed discrepancies, and tracing state evolution across workflow components. We distilled these challenges into three concrete requirements for visual debugging support: cross-artifact evidence alignment, expectation-grounded comparison, and traceable state evolution. Together, these requirements \change{begin to characterize} a design space for future visual debugging research in data-intensive programming.

\section*{Supplemental Materials}
Supplemental materials (interview protocol, codebook, and theme-to-finding mapping) are available on OSF: \url{https://doi.org/10.17605/OSF.IO/KYCWP} (CC BY 4.0).

\acknowledgments{
We used ChatGPT and Claude to assist with language editing, such as improving sentence structure and wording clarity. All research decisions, analyses, interpretations, and writing were conducted and verified by the authors.
}

\bibliographystyle{abbrv-doi-hyperref}
\bibliography{template}
\end{document}